# The Isotopic Field Charge Spin Assumption

**György Darvas**

Symmetrion
and
Institute for Research Organisation of the Hungarian Academy of Sciences
18, Nádor St., Budapest, H-1051 Hungary
darvasg@iif.hu

**Abstract** This paper discusses fundamental physical interactions starting from two preliminary assumptions.
(a) Although *mass of gravity* and *mass of inertia* are *equivalent* quantities in their measured values, they are *qualitatively not identical* physical entities. We will take into consideration this difference in our equations. Then it extends this *'equivalence-is-not-identity' principle* to sources of further fundamental interaction fields, other than gravity.
(b) *Physical interactions occur between these* qualitatively different *entities*.
First it interprets these assumptions. Then, it sketches a picture of fundamental physical fields influenced by the distinction between the two qualitative forms of the individual field-charges and interaction between them. It applies the results of a former publication (Darvas, *Concepts of Physics*, 2009, 3), which mathematically proved the existence of an invariance between the two isotopic forms of field-charges. It introduces the notion of isotopic-field-charge-spin, proven as a conserved quantity. This conservation predicts the existence of a boson mediating between the two possible isotopic-field-charge-spin states. After these preliminary foundations, it formulates certain consequences in the author's view on the physical structure of matter. Finally the paper discusses how these issues can allow an alternative interpretation of physical experience.

## 1 INTRODUCTION

### 1.1 Preliminary Assumptions

The proposed conceptual framework and assumption[1] on the interaction mechanism goes beyond the Standard Model (SM). Many physicists are convinced that SM does not hold eternally alone and is not untranscendable; there will appear new, more precise theories that will partially include the SM, and answer those questions that are left open by the SM. However we do not certainly know how, at least at present.

CERN organised three workshops to discuss possible theoretical candidate models beyond the SM to base a "new physics" in accordance with fine scale anomalies and symmetry breakings in high energy experiments, in 2005-2007 (CERN workshop, 2008a; CERN workshop, 2008b; CERN workshop, 2008c). They agreed that SM holds, it needs only some extensions. So do we as well. Section 4 of the present work provides an alternative extension theory, still not discussed in those three working group reports.

---

[1] The topic of this paper is set forth in more details in the book *Another Version of Facts – On Physical Interactions* by the same author coming soon at the Springer.

This work (started in January 2001) is an attempt to exceed a couple of the limits of the Standard Model. Gerard 't Hooft expressed his view on the physics after the SM: "What is generally expected is either a new symmetry principle or possibly a new regime with an altogether different set of physical fields." ( See in: Hooft, 2005, Sec. 12). The *isotopic field charge spin conservation* and the **D** *field*, being introduced in this paper, are candidates.

The presented idea is based on the same facts like those considered in the SM, only it clusters the observations in another way. Unlike existing alternative theories, e.g., the SUSY, which renders a new ("supersymmetric") brother to each particle, this model clusters the observed sources of fields in two-eggs twin pairs, regarding them as isotopic states of each other, and there is left "only" the twin brothers of the bosons mediating their interactions to be observed. It covers gravitational, electroweak and strong interactions. In contrast to the SUSY, which renders fermion-boson pairs as new-born brothers to each other, the Isotopic Field Charge Spin (IFCS) assumption, proposed in the present work, renders fermion-fermion and boson-boson twins to each other.

This assumption does not assume new *fermions*; the twin brothers of fermions originate in splitting the existing ones. Fermions split as a result of a newly interpreted property. The assumption is mathematically based (Darvas, 2009) on an invariance of interactions under rotation of the isotopic field charges' spin (a property that distinguishes the field charge twins from each other) in a still hypothetical gauge field (called **D**), that means, on the conservation of the isotopic field charge spin.

The *bosonic* twin brothers should appear as the quanta of the **D** field (cf., Section 4.2.3) that mediate between the split fermion states, that means, between isotopic states of field charges. The prediction of bosonic twin brothers will be discussed in Section 4.2.5.

The IFCS assumption theory does not give a clue to everything, (e.g., mass). It is a modest attempt to answer *a few* open questions of contemporary physics.

### 1.2 Identity and equivalence

*Invariance* under a *transformation* means the absence of change (constancy in the *properties* of a given *object*). In respect of the properties of the given 'object', we speak about *identity*, if all of its properties are conserved (unchanged) and indistinguishable; and we speak about *equivalence*, if certain properties of the object are conserved, but not all, and the compared states of the object (before and after the effect of the given transformation) can be distinguished.

In the case of *identity* we speak about 'the same object' before and after the effect of the transformation, which leaves it identical with itself. We say that the object saved its self-identity. This is not so evidently unambiguous in the case of *equivalence*. In certain instances we speak about 'another object', if a transformation changed at least one of its properties (e.g., after charge conjugation), while in others we speak about 'the same object' (e.g., after flipping spin over), which differs from the previous one in one (or a few) of its properties, but its other properties saved their value (i.e., remained equivalent).

Equivalence states something as a result of a comparison between two objects. When we speak of equivalence, we compare non-identical objects, or non-identical (i.e., distinguishable, two) states of an object. We can say, either that two (distinct) objects are equivalent, or that given properties of two objects are equivalent. In both formulations we mean, their given properties are of equal value.

Equivalence presumes the existence of (at least) two objects which can be distinguished in certain properties to make us able to state the characteristic values of their other property/ies to coincide, i.e., being equivalent. In another interpretation we can compare the states of the same (identical) object before and after a symmetry transformation, i.e., when the compared things are different states of the same object observed in different times. (Cf. the definitions of symmetry in Darvas 2001, 2004 and 2007.) In a more strict sense, *identity refers to the 'objects' themselves*, while *equivalence refers to a 'class of properties' of the objects* (Darvas, 2006).

In general, if two physical objects differ in their geometrical or classical mechanical properties, we consider them identical. If they differ also in other properties, we consider them different objects, between which we can ascertain the existence of an equivalence.[2]

In most cases we can avoid ambiguity. In most of the cases we are not concerned about all properties of the compared two objects. We speak about their *equivalence* with respect to a chosen set of properties. Equivalence means that the values assigned to the individual properties in the chosen set are equal. In this case we either neglect or do not know their further properties. Nevertheless, equivalent objects may differ in other properties, which are actually not considered. The consequence is that equivalence and identity can be applied to the same physical objects in different contexts.

In a strict sense, however, *identical objects cannot be equivalent*. Only *qualitatively different* objects can be compared to conclude a *quantitative equivalence* between them. Equivalence always presumes the existence of at least one property, in which the compared objects differ.

## 1.3 The equivalence of the masses of gravity and inertia

The example of the necessity that forced to introduce colours to characterise quarks demonstrates that we can never be sure that there are no further properties in which two objects differ. In other words: equivalence does not guarantee identity.

The equivalence principle is one of the main pillars of the general theory of relativity (GTR). The equivalence principle states the equivalence of the gravitational and inertial masses.

The equivalence principle (at least in its so called Einstein's form, which lays between the weak and strong formulation, but shows more similarity to the weak) says, that a test mass, which is affected by a mechanical force, at least locally, cannot make distinction between the sources of the force [the causes of its acceleration], whether it is caused by the gravity of another mass, or by an inertial force. The equivalence principle states that the inertial mass and the gravitational mass of the test body are proportional, and (as we fixed the factor of proportion to "1") are measured on the same scale. Nevertheless, they should be considered not identical properties. (For the role of the equivalence principle and its extension, see also Mie, von Laue, Einstein, de Broglie, de Haas, Castellani.) As we have shown, identical things cannot be equivalent. Equivalence is a quantitative relation between qualitatively different (non-identical) entities. Only different things can be compared and proven to be equivalent. One needs to have two different qualities to claim they are of equivalent quantities.

---

[2] For example, the distinction between the proton and the neutron exposes an already classical interpretation of the identity-equivalence problem. We quote the example of the isotopic spin because it will play an analogical role in our further discussion in this work. To describe the symmetry between the proton and the neutron, i.e., their interchangability, there was proposed that the proton and the neutron be considered identical particles that can take two states of a property called *isotopic spin*, or shortly isospin. This name expresses that they are two observed isotopic states of an identical particle. The distinction is similar to the spin states of any other particle, which are also considered identical. E.g., two electrons in an 'up' and 'down' spin state are considered electrons both, and not particles denoted by different names. Unlike the spin, which is a property (own angular momentum) described in the space, the analogously introduced isotopic spin property presumes the existence of an abstract (gauge-) field, where it can be described. Nucleons are considered identical particles with the condition, that they differ only in the orientation of the isotopic spin, a property defined in an abstract gauge field. Their identity holds under the condition that they are confined within a nucleus. They lose their identity (i.e., their properties lose their equivalence) as soon as one observes them as free particles outside a nucleus. However, can entities considered to be identical, which are characterised by different quantities of the value of a property?

Physics textbooks conclude from the equivalence principle the identity of the gravitational mass and the inertial mass. We will argue for they are not identical. Nevertheless, they are different properties of the matter. Not only because we denote them by different names. The essence to establish the principle, and to demonstrate experimentally the equivalence, was that we had gained experience about the two mass properties from different observations. The goal of the experiments to demonstrate their equivalence was to construct measurements, where *both* properties are present. We speak about 'both' in the case of qualitatively different entities. The gravitational mass and the inertial mass are qualitatively different entities, which proved to be equivalent in the measure of their effects. They are equivalent in their value (measure of mass), but they are not identical properties.

For school level calculation purposes it was enough to denote both masses with the identical '*m*'. This can be justified, for their values are equal. However, this cannot satisfy our curiosity about certain theoretical (not just mathematical), philosophical and conceptual consequences. Therefore, let's consider the mass of gravity and the mass of inertia as two different properties of matter. For the same massive object can behave once as a source of gravity, then as a measure of inertia, we will imagine them as two isotopic states of the same property, called mass of the object.

## 2 DISTINCTION BETWEEN ISOTOPIC FIELD CHARGES (ך)

As much as the mass is the field charge of the gravitational field, we will call its two isotopic states as *isotopic field charges* (ך) of the *gravitational interaction*. The gravitational mass is associated with the potential part of that interaction, while the inertial part with the kinetic part. In GTR the latter is attributed to the momentum densities, while the former to the gravitational field energy. They are separated within the stress-energy tensor, but according to the general relativity principle they can be transformed into each other – we should add, at least in their quantitatively equivalent values. GTR does not make any statement about the qualitative transformation of the two kinds of masses into each other. This was a reason to identify them. Nevertheless, the distinction cannot be avoided.

### 2.1 Field charges and isotopic field charges

The sources of the individual fields are their charges. The mass of gravity – i.e., the source of the gravitational field – is the field charge of the gravitational field. In a similar way, the electric charge – i.e., the source of the electromagnetic field – is the field charge of the electromagnetic field; and analogously so on, the flavour, lepton charge, for the weak, the colour charge for the strong field. All fields are supposed to have sources. These sources – *field charges* – are assumed to be realised in the matter field, while they serve as sources for gauge fields. Are they really the same, or can one distinguish the two agents?

The mass of gravity and the mass of inertia will be considered as two equivalent quantity *isotopic states* of the field charge of the gravitational field (Isotopic Field Charges). They represent two different qualities. Their concepts express two properties of matter, whose existence originates in different experiences. Physics established quantitative relations between them (i.e., equal values), however this fact does not vanish their qualitative difference. We have all reason to make distinction between them in our theories.

Mass of gravity is a scalar quantity, while the mass of inertia, as a quantity, appears in the three components of a vector, with three (current) components as three projections in three independent spatial directions. In a frame of reference in rest the quantity of the mass of gravity and the mass of inertia are measured equal. According to the equivalence principle their measures are proportional, and the factor of proportion is fixed as „1".

The mass of gravity is associated with the potential part $V$ of the object's Hamiltonian. The mass of inertia is associated with the kinetic part $T$ of the object's Hamiltonian. According to this observation we can call the mass of gravity as (scalar) potential mass, and the mass of inertia as kinetic mass. Consequently there are two qualitatively different proportional factors belonging to the potential and kinetic energies of an individual object, and maintaining that they quantitatively coincide, we should reflect their qualitative difference in our equations of motion.

We can assume that similar distinctions exist for the charges of the other interaction fields. First we extend the equivalence principle for the electric charge, then generalise the principle for any field charge.[3]

## 2.2 Equivalence principle for electric charges

Taking into consideration the covariance of the electromagnetic field, namely, what could look like an electric field for one observer could be a superposition of an electric and a magnetic field for the other.[4] In an idealised case, if the velocity of the moving charges for the latter's reference frame is relativistically high, the effects of the electric field can be neglected compared to that of the magnetic field, since they are magnitudes weaker. The former is similar to the gravitational effect (derived from the potential part of the Hamiltonian), while the latter shows similarity to the inertial effect in mechanics (derived from a kinetic part of the Hamiltonian; cf., the Lorentz force).

A $q$ test charge, which is affected by an electromagnetic force, at least locally, in a small environment, will not be able to make distinction between the sources of the force (the causes of its acceleration), whether they are caused by the electric force of another static charge, or by a magnetic force caused by a high speed moving charge, a current, or a magnetic field of a different source. This is very similar to the equivalence principle formulated above for masses.

---

[3] For example, when we concluded the conservation of the mass solely from the gravitational potential, we ignored any possible contribution by the kinetic part of the Hamiltonian (while the full Hamiltonian was generated by the energy-momentum tensor in the general theory of gravitation). It was similar to that, when we derived the conservation of the electric charge – in classical electrodynamics – from the Maxwell equations alone, we derived an invariance solely from a transformation in the Coulomb field. Thus – in classical electrodynamics – we did not couple it with a transformation in the gauge field (which latter was generated by the rest of the electromagnetic field tensor). This latter „imperfection" has been corrected by the coupled gauge transformation in QED, and a similar „correction" is expected to be done in case of the conservation of mass. (In a proper gauge theory, symmetry transformations leave the total Hamiltonian invariant, and do not the kinetic and the potential components of the energy separately.)

[4] This demonstrative formulation refers to 't Hooft (2002). Nevertheless, the conceptual interpretation of covariance evoked much debates in 20th century physics. Wilczek (1998) noticed, that "when physicists refer to general covariance, they usually mean the form-invariance of physical laws under coordinate transformations following the usual laws of tensor calculus, including the transformation of a given, preferred metric tensor. ... From a purely mathematical point of view one might consider doing without the metric tensor; in that case general covariance becomes essentially the same concept as topological invariance." He also reminds, that in the understanding that Abelian gauge invariance, and in its non-Abelian generalization – as a symmetry under transformations of quantum-mechanical wave functions – "the space-time aspect is lost. The gauge transformations act only on internal variables." Topological aspect is considered in the sense of the generalised variables in the 2nd Noether theorem. In our following treatment space-time co-ordinates take a role of internal variables. For the specification of these internal variables in this generalised sense see the detailed presentation of the proposed theory in (Darvas, 2009) and in the last Section of this paper.

Depending on the chosen system of reference, in accordance with the covariance principle, the same charge can be in rest or moving with a high velocity. Therefore the same object can behave once as the source of a Coulomb force, and in another frame of reference as a source of a (kinetic, Lorentz) magnetic force. We can consider this so, that the same agent can be either a source of a Coulomb force, or the source of a magnetic force, and these are two properties of the given charge. In other words, they have two states of the same property (i.e., of the charge) of the object.

The *equivalence principle for electric charges* means, that charges in the two states are equivalent in their charge value, but they are not considered identical physical properties. They are two, different qualities, and the same object can occupy any of the two, isotopic, states, namely a potential (Coulomb) charge, and a kinetic, current-like (Lorentz) charge. Since the electric charge is the field charge of the electromagnetic field, they are two *isotopic states* of this field charge. Note, both negative and positive electric charges can occupy both isotopic states, respectively. The Coulomb charge is a scalar property, while the current-like charges have three components of a vector. The four charge components together form a four-current.

Looking for analogies, note also, that the gravitational mass (as an isotopic gravitational field charge), and the Coulomb charge (as an isotopic electromagnetic field charge) are associated with the scalar potentials of the given fields.

The charges composing electric current can be associated with the vector potentials of the electromagnetic (gauge) field, and similar can be assumed on the inertial mass in a gravitational field, and these latter do not serve as sources of the (scalar) electric and the gravitational fields, respectively.

**2.3 Distinction between isotopic field charges of the electromagnetic field**

We can extend our above separation program (formulated for the charges of the gravitational field) to the charges of the electromagnetic field. This means, we can replace the charges appearing in our equations by two different (isotopic) charges, a Coulomb-type one, and a kinetic-type one. We do it in a similar way, like we did in the case of electromagnetic currents, only now we extend the distinction with the introduction of the two isotopic charges.

In classical electrodynamics, the $A_\mu$ four-potentials of the electromagnetic field were invariant under Lorentz transformation and the four-current $j_\nu$ components transformed like a vector. We assumed, that the sources of the Coulomb force ($q_C$) are different type charges, than moving charges as sources of currents ($q_T$ type charges, where the index $T$ refers to their kinetic character). The same charges play both roles (cf. covariance), we assume only, that in the two situations they behave as two isotopic states of the same physical property (i.e., field-charge). Provided, that the fourth component (in + + + – signature) of $j_\nu$ current density, namely $j_4 = ic\rho$ contains a different kind of charge-density ($\rho_C$), than those moving (current-like, kinetic) charges ($\rho_T$) in $j_i$ ($i$ = 1, 2, 3), the **j** current would lose its invariance under Lorentz transformation. Since there appear different type (isotopic) charges in the individual components of this new $j_\nu$ current density, this latter $j_\nu$ does not transform like a vector. It is easy to demonstrate that this $j_\nu$ is *not* a four-vector, and for the $\rho_C$ and $\rho_T$ are mixed during the multiplication by $F^{\mu\nu}$, the components $F^\mu$ do not transform as vector components either. The result of this example is not in compliance with our experience! We lost some symmetry. As a consequence, to restore Lorentz invariance and compliance with experience, we must require the existence of an additional transformation that should counteract the loss of symmetry caused by the introduction of two isotopic states of the charges.

A similar consideration can be formulated for the inertial and gravitational masses, and the four-potential of the gravitational field. Of course, the concrete forms will be different for the gravitational field and the electromagnetic field, due to the difference between the energy-momentum tensor and the electromagnetic field tensor. Nevertheless there are certain similarities as concerns the application of the equivalence principle for the four-currents in both cases. Identifying this invariance transformation, in accordance with the general covariance principle of nature, is the subject of the rest of this paper. One of our main results in this work (mathematically proven in Darvas, 2009) is the derivation of the missing transformation that counteracts, and replaces in its role, the demonstrated apparently lost symmetry.

## 2.4 Isotopic field charges in strong and electroweak interactions

Certain similitude between the relation and different physical roles of gravitational mass and inertial mass on the one hand, and of Coulomb charges and kinetic charges on the other, seems obvious. They are field charges of different kind fields, however, the gravitational mass and the electric Coulomb charge serve both as sources of their respective fields, associated with a central, scalar potential, and represent bound states of the respective field charges. At the same time the inertial mass and the kinetic electric charge are both associated with the kinetic part of the Hamiltonian of their interaction field, both appear in vector components of currents and are the sources of momentum-like forces, namely of the angular momentum and the magnetic momentum, and both represent free states of the respective field charges.

We cannot observe the same doublets in the case of colours, the charges and sources of the strong field. Since quarks, the carriers of the colours are confined in the baryons, it is hard to detect their free states, and we can conclude the existence of such states only with the assumption of the asymptotic freedom. We may assume that if one could detect unconfined quarks, one would observe similar bound and free, potential and kinetic, scalar and vector property couples in their colour charge behaviour. Hard radiation of quarks, what is a rare phenomenon compared to the soft radiation, can change the overall flow of the energy and momentum in a large extent (Wilczek, 2003). This situation corresponds to an (asymptotically) free state of fast moving quarks, which represent current-like, kinetic states of the carrier of the strong field charge. However, confined quarks cannot move very fast for long due to the limited diameter of a baryon. So their interactions involve soft radiation, which is detected in majority of cases. The more common soft radiation allows to conclude to the bound states of the carriers of the colours, which should be scalars and represent the sources of the strong field. Considering hard and soft radiation, we assume that (all the three) colour charges appear in two isotopic states as well.

Electroweak interaction is subject of combined [$U(1) \times SU(2)$] symmetries. Conservation of several charges are associated with the electroweak interaction, like that of the weak hypercharge, weak isospin, lepton number, baryon number, and the conservation of quark flavours (Glashow and Weinberg, 1977) in weak neutral current effects in generalized Weinberg-Salam models. The material carriers (potential sources) of these electroweak interactions are baryons and leptons, characterised by the flavours, electric charges; baryon numbers of the carriers of the flavours, and lepton numbers of the carriers of electric charges, as well as the combined weak hypercharge. They appear in their kinetic form in the so called ***B*** (for $U(1)$ symmetric) and ***W*** (for an $SU(2)_L$ symmetric) fields of the electroweak interaction's Lagrangian.

## 2.5 Preliminary assumption of a gauge transformation between isotopic field charge states

At least, our experience obtained from masses and electric charges, weak charges, and the indirect observations on strong colours allow us to assume a field-independent property – what we called isotopic field charge – which is responsible for this similar, double behaviour. We assume isotopic field charge to be a property that can be attributed to field charges of any known interaction field.

To justify the meaning of the obtained results, the two kinds of charges of the electromagnetic field and the two kinds of charges of the gravitational field, field charges of the electroweak interaction, and presumably of the strong field, respectively, should be transformed into each other by a gauge transformation. Such a gauge transformation should involve the existence of a conserved property that we define in the following way.

Since the required transformation affects the *isotopic state*s of the individual *field-charges* (ד), this transformation must be performed in a special gauge field; and since these states can occupy two positions in that gauge field, it must be a *spin-like property*, therefore, we will call this property as *Isotopic Field-Charge Spin (IFCS)* and denote it by $\varDelta$, and we will refer to the invariance transformation what we are seeking for as *isotopic field-charge gauge transformation*. This assumption assumes the existence of a local gauge field, in which the isotopic field-charge spin can rotate and occupy two states and concludes a conserved (non-Abelian) current and a corresponding class of *SU*(2) type invariances.

For the same object can behave, e.g., in the electromagnetic field, once as the source of a Coulomb force, and in another frame of reference as a source of a (kinetic) magnetic Lorentz force (cf., covariance principle), they must be able to get transformed into each other. Non-Abelian character and arbitrariness involve that the orientation of the isotopic field-charge spin is of no physical significance. If we determine the proper form of this invariance transformation, it will counteract the loss of symmetry between the two kinds of field-charges, and bring our equations in compliance with the experimental results.

The required invariance shows certain formal similarities to YM-type invariances (1954). However, it must differ from them in at least two features. Once, the concerned physical property, namely the isotopic field charge (IFC, ד ['dalet' the fourth letter of the Hebrew alphabet]), is a quite different physical property than the isotopic states of nucleons. Secondly, the gauge field, and consequently the gauge transformation that rotates the isotopic field charge spin (IFCS, $\varDelta$) in this gauge field, are quite different from the isotopic gauge field derived for the isotopic spin transformation. (For specification, see the subsections 4.2 of this paper.)

The existence of such an invariance transformation will provide us with a symmetry, and consequently with a conservation law, namely the conservation of the introduced new property ($\varDelta$) of the field-charges. The conservation of isotopic field-charge spin is identical with the requirement of invariance of all interactions under isotopic field-charge spin rotation (in the gauge field where it is interpreted). This means, that all physical interactions should be invariant under a transformation in an isotopic gauge field, more precisely, under a rotation of the property, called isotopic field-charge spin ($\varDelta$). We will show that there exists such an invariance transformation.

The IFCS (isotopic field-charge spin, $\varDelta$) is attributed to the field-charges as entities, independent of which physical interaction field's charges are considered. Its assumption introduces not only a *property* that differs from all described ones, but also a *field* in which it can be interpreted.

Concluding the existence of a conserved property called IFCS ($\Delta$) assumes a gauge field that can be introduced partly analogously to the YM field, however, it differs from that, for there is a quite different (i.e., different from the isospin) property that should be rotated and taking two stable orientations in that (different) gauge field. We will derive this field and the conservation of IFCS below in section 4.2.

## 3 INTERACTION BETWEEN ISOTOPIC FIELD CHARGES (ד)

The second preliminary assumption of this book is to conjecture interaction between isotopic field-charges.

In the years of the birth of QED, Dirac (1928, 1929), then also Fermi (19391, 1932), Breit (1929, 1932) started from a picture in which they supposed that there are the (static, scalar) Coulomb potentials of the electric charges what initially interact. Then the Coulomb charges bring each other in motion and that propagates the vector potentials and interaction of those too. E.g., Dirac applied radiation gauge, and in zero approximation the unperturbed Hamiltonian contained the Coulomb interaction only, the next parts of the perturbation introduced the effects of the vector potential. Heisenberg supported the same model of perturbation.

C. Møller (1931) proposed an opposite way. He considered a kinetic interaction in the unperturbed Hamiltonian, and took into account the Coulomb interaction in the perturbation. His approach was not widely accepted, for it included an *asymmetry in the role of the interacting charges*.

In 1932, Bethe proposed Fermi to describe the interaction between two charged particles assuming that they were initially in free-particle states both (and to apply Lorentz gauge), so that they modified the model proposed by Møller. They introduced a(n artificial) *symmetrisation* in the roles of the interacting particles (Bethe-Fermi, 1932). However, their main conclusion concentrated not on the unperturbed kinetic model, rather to prove that the model proposed by Møller is an equivalent description with the former Dirac-Fermi-Breit type description of the interaction between two electrons. In this picture (following Møller), the interaction of two approaching particles starts from the infinity, where the Coulomb fields affecting each other are weak, and thus, can be neglected. They used perturbation theory too, where the scalar potential appears in the next approximations when the interacting particles get closer to each other in a process of scattering. The symmetrisation used by Bethe and Fermi – and which was introduced to the equation artificially – fitted better in the picture about the mechanism of interactions prevailing in the early thirties, although it was shown later that their calculations lead to deviations from the experimental data (Araki and Huzinaga, 1951; Salpeter and Bethe, 1951).

In short, we had two models in classical QED. One, in which the interaction starts between two bound states of particles, and another, in which the interaction starts between two free state particles. In the first model the effect of the vector potential enters only in the perturbation, in the second model, the role of the scalar potential is left for the perturbation. The main conclusion of the Bethe-Fermi paper (1932) was that the two descriptions are equivalent. (More details about the birth of these different approaches to QED see in Schweber, 2002; and Dyson, 2005.) Both pictures are approximations, and both lead to relatively good results in accordance with the experience.

This paper proposes an *intermediate model*, in which the interaction takes place between bound and free state particles initially. Please, note, if the model of the initial interaction between two bound states, and the model of the initial interaction between two free states could have been considered equivalent (Bethe and Fermi, 1932), there can be found an equivalent place among the physical descriptions for the intermediate model as well. However, such a model has not been described.

Nevertheless, the intermediate model allows a further interpretation: exchange between the two isotopic states of a single particle.

**3.1 Single particle's isotopic field charge states**

To understand the possible mechanism of the intermediate model of interactions, first we discuss the change in isotopic states of field charges first on single particles.

When we take a measure on an object, we have no experience that we found it in one or the opposite isotopic state. When we observe a single particle, it will be either in one or in the other IFCS state. We can call the two states as potential and kinetic, scalar and vector, or bound and free states. However, our measurement records a mixture of the two states. Nevertheless, we do not observe the individual IFCS states. Our observation suggests that they behave as being in both states, each measured object can occupy both a potential (bound) and a kinetic (free) IFCS state. In the lack of experience to catch a particle in one or the other stable state, we have good reason to assume that they permanently change their states. (Randomly or with a stable frequency, we do not know, they may probably follow a similar mechanism like quarks do during their colour change via gluon exchange).

Let us consider a model of a doublet, when a particle can be in a potential state ($V$) and in a kinetic state ($T$). According to its actual state it has potential or kinetic energy respectively. According to our observation all particles possess both. We can interpret the phenomenon in the following four ways: probabilistic model, harmonic oscillator model, flip-flop model, intermediate particle model.

*1* In *the probabilistic model* we can consider that the wave function of the given particle may be in a potential state with amplitude $\psi_V$, or in a kinetic state with amplitude $\psi_T$. The wave function of the given single particle is a mixture of the probabilities being in the potential or the kinetic state:

$$\psi = \begin{pmatrix} \psi_T \\ \psi_V \end{pmatrix}.$$

We detect this probabilistic mixture in a measurement. In a large set of particles (e.g., in the case of a massive body consisting of many particles) the probabilities reach a stable proportion and we observe stabilised measurable potential and kinetic energies in a given reference frame.

*2* *The harmonic oscillator model* presumes that – in the simplest case and in first approximation – the Hamiltonian of the particle consists of $V$ and $T$ only. The model assumes that the energy of the single particle is concentrated either fully in $V$ or fully in $T$ and the energy commutes periodically between the two states. The transition is assumed to follow a harmonic function in time. We can take a measurement of the energies in a transitional phase, which can be interpreted as a mixture of probabilities being in the two extreme states, similar to the wave function in the probabilistic model. However, in the case of a single particle, physical meaning can be attributed only to the two extreme states. As mentioned, in this model the energy of the particle is concentrated either in $V$ or in $T$. This has certain consequences. When our physical object (i.e., field charge, denoted by ד [dalet]) is in the *potential state* – in our notation – ד$_V$ *(its full energy is V)* its features show up as a *scalar particle* and it behaves as a *source of a scalar field*. When it is in the *kinetic state* – in our notation – ד$_T$ *(its full energy is T)* it behaves as a

*wave,* its field charge current takes the form of a *vector,* and *generates a vector field* around itself. Unlike the field charge of the scalar potential field, whose quantity is invariant (bare or rest field charge), the quantity of the field charge (more precisely, the three independent components of its current) in kinetic state varies according to its velocity in the respective reference frame in which it is actually observed (according to the Lorentz transformation and in accordance with the general covariance principle). The harmonic oscillator model presumes the permanent change of a single particle between its two isotopic field charge states along with the changes in the listed consequences.

3  *The flip-flop model* differs from the harmonic oscillator model that the change between the two isotopic field charge states is assumed to undergo suddenly, without transition. This model considers that the Isotopic Field Charge Spin ($\varDelta$) can rotate in a presumed **D** field not smoothly, rather it can take only two quantum positions. (See the mathematical foundations in the cited paper by the author, and note that the mathematical foundations were developed in advance to the phenomenological model [Darvas, 2009]. There are didactical reasons for we are presenting here first the phenomenological description of the physical model.)

4  *The intermediate particle model* presumes that the switch between the two states happens with the mediation of a boson that governs the transition from one isotopic field charge state to the other. This boson may not carry any physical quantum number but the $\varDelta$. Therefore it cannot be identical with any existing known intermediate boson. It should act parallel (or antiparallel) with them. This is why we assume it as their (two eggs) twin brother boson.

The common feature in the four models is that any single particle carrying a charge of a fundamental physical interaction field is in permanent change between the two isotopic field charge states. Since it is difficult to observe them in one or the other isotopic field charge state we attribute both properties to each of them. This is the background of the long rooted *corpuscle-wave double behaviour* discussion.

As mentioned, a particle *in a potential state* behaves like a *corpuscle*, while *in a kinetic state* it behaves like a *wave*. Since the two IFC states switch between each other with a high oscillation frequency one cannot detect a single particle in a full corpuscle or in a full wave state. What we can detect that depends on the nature of the actual interaction of the individual particle on the one hand, and we can observe a probabilistic mixture of the two states on the other hand.

A particle *in a potential state* plays the role of the *source of a scalar field*. Therefore a potential isotopic field charge (we will denote by ℸ$_V$) is a scalar quantity. A particle *in a kinetic state* serves as a *current source of a vector field*. So a kinetic isotopic field charge (we will denote by ℸ$_T$) will play role in three vector components of a quantity according to three, directed, independent components of a given field charge current.

## 3.2 The intermediate model of interaction between two particles

We cannot reconstruct, whether Møller (1931) was aware of the same mechanism in 1931 that we are proposing now. Probably, he was not. Nevertheless his original formulation, what he derived, contained a similar asymmetry between the roles of the interacting particles what we are proposing. Bethe and Fermi (1932) considered this asymmetry an incompleteness of the theory by Møller and rectified it. Instead of their artificial correction in the scattering matrix elements we attempt to interpret the original equation of Møller, and fit it in an alternative picture.

Bethe and Fermi (with reference to Møller) followed the next logic. In contrast to the derivation by Breit, in Møller's theory the Coulomb energy is a part of the perturbation and enters additionally on the side of the interaction with the (electromagnetic) tension field that was considered initially in the unperturbed approximation (Bethe and Fermi, 1932, p. 20), (i.e., in zero approximation the two electrons behave as independent from each other). They build the retarded potentials of the electrons, where the field of the first electron influences the second electron as a perturbing effect. Then they calculate the matrix elements of the interaction energy of the exchange between two states of the first electron ($n_1$, $n_1$') and two other states of the concerned second electron ($n_2$, $n_2$') resulted in the interaction, while the sum of their energies is conserved. The calculated matrix element describing the transition of the second electron from state $n_2$ to state $n_2$' is:

$$V_{n_1 n_2}^{n_1' n_2'} = e_1 e_2 e^{\frac{2\pi i}{h}(E_1' + E_2' - E_1 - E_2)t} \int u_2'^*(\vec{r}_2) u_1'^*(\vec{r}_1) \left[ \frac{1}{|\vec{r}_2 - \vec{r}_1|} - \frac{2\pi^2}{h^2 c^2}(E_1 - E_1')^2 |\vec{r}_2 - \vec{r}_1| - \frac{(\gamma_1 \gamma_2)}{|\vec{r}_2 - \vec{r}_1|} \right] u_2(\vec{r}_2) u_1(\vec{r}_1) d\tau_1 d\tau_2$$

The first expression in the bracket [ ] is the corresponding Coulomb potential, the second originates in the retardation of the scalar potential, and the third is the effect of the (unretarded in first approximation) vector potential. Bethe and Fermi found that the first and the third expressions were symmetric for the two electrons but the second was asymmetric. They ascertained that this asymmetry was a result of an incompleteness of the method applied by Møller. Therefore, with reference to the requirement that the full energy in the starting and final states should be equal, they artificially made that expression symmetric and replaced ($E_1$-$E_2$')² by –($E_1$–$E_1$')($E_2$–$E_2$') .

Without continuing their clue, in my opinion this artificial symmetrisation is nice, but it is misleading and physically not well founded. Bethe and Fermi forced an assumption upon Møller, what did not constitute a part of Møller's theory, namely that the two interacting particles must play equal roles. Starting from this point of the derivation by Bethe and Fermi, we must separate their clue, from the clue of Møller. Bethe and Fermi insist on the symmetric role of two interacting particles, while Møller foresaw (or intuitively felt) that two interacting particles could be in two different states, although he could not clearly identify the essence of the distinction between the roles of the interacting agents. In my opinion, we do not need to demand that the individual interaction potentials be symmetric in respect of the two interacting agents. (I assume this claim is in accordance with those put down by Møller, although we are unable to check whether really this was his conscious intention.) We need to demand only, that in an opposite situation – that means, when the second particle plays the active role and the former is the passive (this asymmetry can be exemplified by the emission and absorption of a photon), or mirror scattering – similar (symmetric) potentials be valid and their numeric values coincide with those in the first situation.

This intermediate model can be interpreted in different ways. Remaining at the example of QED, we can interpret that in an initial state the Coulomb potential of a particle interacts with the vector potential of another. A more general possible formulation of the interpretation is that the potential (scalar) part of a Hamiltonian interacts with a kinetic (vector) part. We can say also, that the bound state of a particle interacts with the free state of another particle (what sounds strange in this form, since as soon as the latter particle gets subject of an interaction, it will be free no more, but a first approximation). Finally, using the terminology introduced in this paper, the interaction can take place between particles in two opposite isotopic field-charge states. The latter three interpretations do not reduce their relevance to the electromagnetic interaction.

All these three models (the Dirac-Breit-Fermi model in accordance with the Heisenberg-Pauli formalism on the one side, the Møller-Bethe-Fermi model on the other, and finally our intermediate model based on Møller's presumed original intention) are interpreted for the interaction between two particles. Nevertheless, the intermediate model allows a further interpretation: exchange between the two isotopic states of a single particle.

Summarising the proposed mechanism of the interaction between two particles: (1) first, we assume that the single particles oscillate between two isotopic field charge (IFC, i.e., ꓶ) states – as we saw the possible mechanisms above – that means, they are at any moment either in a potential or in a kinetic state; (2) at second, we assume that the interaction takes place always between a particle in one of the ꓶ states and another particle in the opposite ꓶ state; (3) at third, the two particles switch their ꓶ states simultaneously during the interaction so that both are in an opposite ꓶ state at a given moment.[5] The reality of this presumed phenomenological mechanism will be demonstrated in an exact way in Section 4.

# 4 CONSERVATION OF THE ISOTOPIC FIELD CHARGE SPIN ($\Delta$)

Distorted symmetry of our equations[6] – what is not in accordance with the experienced facts – can be restored by proving that there exists an invariance between the twin brothers of the field charges (sources of the fields) split according to the introduced new property ($\Delta$). Invariance means that particles, disposed with these properties, can be exchanged. The "exchange rate" (gauge) depends on the velocity of the kinetic field charge compared to the respective matter field (i.e., to the potential

---

[5] We must remark, that the problems of the consideration of asymmetries between interacting particles and antisymmetrisation as a method was not far from Bethe when he proposed to Fermi to symmetrise Møller's scattering matrix when studying transitions of interacting electrons between two states in 1932. The method appears in his *Ansatz*, written in the previous year during his first visit in Rome, in which he already expressed his thanks to Fermi for those advises. The Bethe *Ansatz* concerned the interaction of two electrons being in opposite spin positions that respect the Pauli exclusion principle. Seemingly Bethe and Fermi did not accept similar asymmetric roles (of another property) in the interacting electrons when applying the Møller scattering model, and insisted on symmetry.

The symmetric interchangeability of the interacting fields used to be a strong paradigm. It prevailed physical thinking for long, even in the recent decades. Not only Bethe and Fermi (1932) insisted on the symmetry in the interacting physical agents. Note, that much later, Goldstone (1961) assumes also a symmetric role between the interacting scalar fields. P. Higgs (1964) "breaks" this symmetry in the roles of the interacting scalar fields using a seemingly similar consideration like Møller (1931). Our approach differs from Higgs' in the assumption of a velocity dependence in the case of the vector field. S. Weinberg (1967) attributes the asymmetry between interacting agents to chiral configurations. In our 'covariant interpretation' the interacting agents must be in different physical states, but their roles must be exchangeable. The latter can be interpreted in terms like Yang and Mills formulated, only we extend their terms of 'arbitrary choice' of one or the other state of the interacting particles in an *isotopic field* → to the *isotopic field charge field*.

What we do here is to apply the idea of asymmetry to the interaction of particles that must be in opposite isotopic field charge states, so that we acknowledge the asymmetric roles of the agents entering into interaction with each other. In analogy to the Pauli principle formulated for spin, we assume that interacting particles (not only electrons and not only electrically charged ones) should be in opposite IFC states; in other words, they must owe opposite position *isotopic filed charge spin* (*IFCS,* i.e., $\Delta$). Since $\Delta$ is interpreted in a gauge field (what we will denote by *D* [note, that the electromagnetic field was denoted by *A*, the YM isotopic field by *B*, and *C* was occupied by several other purposes in physics])**,** and it concerns field charges of any fields, we have some reason to conjecture that the assumed principle of interaction between opposite IFCS particles can be extended to any interaction field. Thus, *IFCS may play an integrating role among the fundamental physical interactions*. The optimism is cherished by the successful extension of the Bethe *Ansatz* to several areas of quantum field theories, covering all physical interactions respectively, at accelerated pace in the recent decade.

[6] According to P. Higgs (1966): "The idea that the apparently approximate nature of the internal symmetries of elementary-particle physics is the result of asymmetries in the stable solutions of exactly symmetric dynamical equations, rather than an indication of asymmetry in the equations themselves, is an attractive one." Please, compare this notice with Wigner's (1984) concern discussed in Section III. 7.4!

field charge in rest in that field). The validity of the assumption can be verified by demonstrating the existence of the gauge bosons that mediate the exchange.

We know certain phenomena in classical physics that depend on velocity in a given reference frame. In general, kinetic quantities depend first on velocity in the chosen reference frame, and only indirectly, through $v = v(x_i, t)$ on the space-time variables. As Mills (1989) observed, "Hamilton's principle was first discovered in connection with mechanical systems, where the Lagrangian turns out to be the difference between the kinetic and potential energies, but the principle is easily extended to include velocity-dependent forces of certain types", including, e.g., the magnetic force on a moving, electrically charged particle. It is not surprising that phenomena related solely to the kinetic part of the Hamiltonian ($T$) can be described in a velocity dependent, i.e., kinetic field $D_T = D[v(x_i, t)]$ where the dependence on the local co-ordinates is indirect. We refer to (Norton, 2003, pp. 118-120) according to whom "active general covariance allows the generation of the field $\varphi'(x^i)$ from $\varphi(x^i)$ by the transformation $x^i$ to $x'^i$" and the same applies, when we use parameters $\varphi(\dot{x}_\mu(x_\nu))$ (and even in more general cases). This does not disclose the possibility of localisation of the theory in space-time, however, it does not ensure it automatically. Local symmetry in a kinetic field means that the objects, fields or physical laws in question are invariant under a local transformation, namely under a set of continuously infinite number of separate transformations with an arbitrarily different one at every velocity in the given reference frame.

The isotopic field charge (ק) can be identified in the case of the gravitational field with the properties of the masses of gravity and inertia respectively, in the case of the electromagnetic field with the properties of the Coulomb, and current- (or Lorentz) like charges respectively, and so on. The potential isotope of ק (ק$_V$) depends directly on space-time co-ordinates. The physical state of the kinetic isotope of ק (ק$_T$) depends primarily on the components of its velocity (and indirectly on its space-time co-ordinates). When we try to specify physical phenomena that distinguish kinetic behaviour of objects from their behaviour in a field caused by another, potential source (i.e., ק$_V$) we should make attempt to seek for a description in a velocity dependent field.

### 4.1 Mathematical background

(Darvas, 2009) was seeking for invariance between scalar fields and (gauge) vector fields that describe kinetic processes, the latter depending therefore primarily on velocity. For this reason, that paper considered Lagrangians which depended on matter fields $\varphi_k$, and gauge fields $D_{\mu,\alpha}$, which all depended – in simple mathematical terms – on $x_\mu$, given in this specific case by the formula:

$D_{\dot\mu} = D_\mu(\frac{\partial x^\mu}{\partial x_4})$, or in another form $D_{\dot\mu} = D_\mu\left[\dot{x}^\mu(x_\nu)\right]$, where $\dot{x}^\mu = \dot{x}^\mu(x_\nu)$; ($\mu, \nu = 1, 2, 3, 4$);

(dotted indices denote the velocity-space components), with a notation $\lambda_\mu^\nu = \partial_\mu \dot{x}^\nu = \frac{\partial \dot{x}^\nu}{\partial x_\mu}$ (Lorentz invariant acceleration), which characterised the changes of the velocity-space components in the space-time. *Localisation* of **D** was taken into consideration in this way $D_\mu\left[\dot{x}^\mu(x_\nu)\right]$.[7] This choice was in full accordance with the conditions set for Lagrangians in the 2nd theorem of Noether.[8] The proof discussed general, non-Abelian case. The mathematical derivation based on a transformation group $G$ and the transformations of its elements into each other $T\left[G_{\infty,\rho}\right] = T\left[p_\alpha(x_\beta)\right]$, where the number of parameters were arbitrary finite numbers $(\alpha = 1,...,\rho)$; , $(\beta = 1,...,\sigma)$. The $p$ were parameters on which the transformations, constituting the group elements, depended. They took the form of functions $p_\alpha(x_\beta)$ and their derivatives. The group transformations depended on $p$ and were finitely differentiable. $G$ may take the form of different groups, depending on the concrete form of interaction in subject, namely $SO(3,1)$, $U(1)$, $SU(2)$, $SU(3)$ in the case of the fundamental physical interactions.

---

[7] Relativistic covariance under a Lorentz transformation $S(\Lambda)$ and its consequences are a standard part of quantum field theory textbooks for long, e.g., Itsykson and Zuber, 1980, Sec. 2.1.3. Here we take into account time derivatives of Lorentz transformed velocities.

[8] In mathematical terms she did not specify either the physical-mathematical character or the number of applicable parameters.

The considered Lagrangian density was $L(\varphi_k, D_{\mu,\alpha})$, where $\varphi_k$ ($k = 1, ..., n$) were the matter fields – which also included the velocity field $\dot{x}^\mu = \dot{x}^\mu(x_\nu)$ –, and $D_{\mu,\alpha}$, ($\alpha = 1,...,N$), were the (kinetic) gauge fields. It was assumed, that $L(\varphi_k, D_{\mu,\alpha})$ was invariant under the local transformations of a compact, simple Lie group $G$ generated by $T_\alpha$, ($\alpha = 1,...,N$), where $\left[T_\alpha, T_\beta\right] = iC_{\alpha\beta}^\gamma T_\gamma$, and $C_{\alpha\beta}^\gamma$ are the so-called structure constants, corresponding to the actually considered individual physical interaction's symmetry group.[9,10] Finally, it considered a local transformation $V(\dot{x}) \in G$ parameterised by $p_\alpha(\dot{x})$ that acts on $\psi$ as $\psi = V\psi'$ in the form of $V(\dot{x}) = e^{-ip_\alpha(\dot{x})T_\alpha}$, which was localised by $V = V\left[\dot{x}^\mu(x_\nu)\right]$.

With the listed conditions, (Darvas, 2009) demonstrated that in the presence of a gauge field **D** there appear two (families of) conserved Noether currents

$$J_\alpha^{(1)\nu} = \partial_\mu F_\alpha^{(1)\mu\nu} \qquad\qquad \partial_\nu J_\alpha^{(1)\nu} = 0 \qquad\qquad (1)$$

$$J_\alpha^{(2)\nu} = \partial_\mu F_\alpha^{(2)\mu\nu} \qquad\qquad \partial_\nu J_\alpha^{(2)\nu} = 0 \qquad\qquad (2).$$

This set of equations should be completed with:

$$\frac{\partial L}{\partial(\partial_\mu D_{\dot{\nu},\alpha})}\lambda_\nu^\rho + \frac{\partial L}{\partial(\partial_\nu D_{\dot{\mu},\alpha})}\lambda_\mu^\rho = 0 \qquad\qquad (3)$$

Although the two conserved currents are not independent, in the presence of a kinetic (velocity-dependent) gauge field they exist simultaneously. ($\lambda_\mu^\nu$ mixes the components of the gauge-field currents $J_\alpha^{(1)\mu}$ depending on the 4D velocity space in a similar way, like the Lorentz transformation mixes the co-ordinates of four-vectors in the 4D space-time; since the $\lambda_\mu^\nu$ tensor was defined to characterise the changes of the velocity-space components in the space-time.)

According to (Darvas, 2009), one can now write $J_\alpha^{(1)\mu}$ as

$$J_\alpha^{(1)\nu}(\dot{x}) = i\lambda \frac{\partial L}{\partial(\partial_\nu \varphi_k)}(T_\alpha)_{kl}\varphi_l(\dot{x}) \qquad\qquad (4)$$

where $\lambda$ denotes a general coupling constant, and $\varphi_l(\dot{x})$ are the matter fields in the Lagrangian density $L(\varphi_k, D_{\mu,\alpha})$. This form coincides with the usual conserved Noether currents known in field theories.

---

[9] (Darvas, 2009) partly followed the clues by Higgs (1966) and Weinberg (1972) at the beginning of their papers with the exception that it considers different dependencies in the potential and kinetic Hamiltonian terms.

[10] For examples, in the case of $SU(2)$ symmetry, $G$ consists of $2 \times 2$ matrices with 3 independent components, representing a state doublet, and in the case of $SU(3)$ its matrix has 8 independent components, representing a state triplet. For simplicity we assume that the matter fields belong to a single, $n$-dimensional representation of $G$.

The most significant conclusion of the above cited derivation is that in the presence of a kinetic (velocity-dependent) gauge field **D**, there appear additional $J_\alpha^{(2)\nu}$ conserved currents in the form

$$J_\alpha^{(2)\nu}(x) = i\lambda \left[ \frac{\partial L}{\partial(\partial_\mu \varphi_k)} (T_\alpha)_{kl} \varphi_l(\dot{x}) \lambda_\mu^\nu - C_{\alpha\beta}^\gamma D_{\dot{\omega},\beta}(\dot{x}) \lambda_\mu^\omega \times F_\gamma^{(2)\mu\nu}(x) \right] \quad (5)$$

where $F_\alpha^{(2)\mu\nu}(x)$ is:

$$F_\alpha^{(2)\mu\nu}(x) = \frac{\partial D_{\dot{\rho},\alpha} \lambda_\mu^\rho}{\partial x_\nu} - \frac{\partial D_{\dot{\sigma},\alpha} \lambda_\nu^\sigma}{\partial x_\mu} - i\lambda C_{\alpha\beta}^\gamma D_{\dot{\rho},\beta} \lambda_\mu^\rho D_{\dot{\sigma},\gamma} \lambda_\nu^\sigma \quad (6)$$

One can observe that the dependence of (4) and (5) on the velocity-space gauge is apparent, although, none of the conserved vector currents involve the gauge parameters $p_\alpha(\dot{x})$.

The cited derivation also obtained that

$$\partial_\mu F_\alpha^{(1)\mu\nu}(\dot{x}) = i\lambda \frac{\partial L}{\partial(\partial_\nu \varphi_k)} (T_\alpha)_{kl} \varphi_l(\dot{x}) \quad (7)$$

and

$$\hat{\partial}_\mu F_\alpha^{(2)\mu\nu}(x) = i\lambda \frac{\partial L}{\partial(\partial_\mu \varphi_k)} (T_\alpha)_{kl} \varphi_l(\dot{x}) \partial_\mu \dot{x}^\nu \quad (8)$$

where the careted $\hat{\partial}_\mu$ denotes covariant derivative[11].

**4.2 Physical discussion of the mathematical foundations**

Relations (7) and (8) provide the equations of motion for the potential part[12] of the system's Lagrangian density. As mentioned in (Darvas, 2009), it is generally the case that when (7) or (8) is satisfied, the matter-field current associated with the Lagrangian acts as the source for the gauge fields. This is a consequence of the fact that the matter-field dependent and the gauge-field dependent currents are at separate sides in each of the latter two equations.[13]

The covariant dependence on the velocity-space gauge field is obvious from (8), and it was shown in a similar way for (7) as $J_\alpha^{(1)\nu}(\dot{x}) = = \hat{\partial}_\mu F_\alpha^{(1)\mu\nu}(\dot{x}) - i\lambda C_{\alpha\beta}^\gamma D_{\dot{\mu},\beta}(\dot{x}) \times F_\gamma^{(1)\mu\nu}(\dot{x})$. *The derived conserved currents make a correspondence between the matter fields and the kinetic (velocity-dependent) gauge fields. They open the way to conclude an invariance between the sources of the scalar fields on the one side, and the gauge vector fields on the other.*

---

[11] Note the following: The YM theory (Yang and Mills, 1954; Mills, 1989) introduced the covariant form of $F_\alpha^{(2)\mu\nu}$ derived from the Lagrangian density of a specific fermion field. We do not make any preliminary assumption concerning the Lagrangian density of the field. (Darvas, 2009) defined the covariant $F_\alpha^{(2)\mu\nu}$ in an essentially different, independent way, based on the requirement of their invariant transformation, and thus got rid of any specific form of the Lagrangian density. The importance of this different approach becomes apparent looking at the discussion of the results by R. Mills himself (1989) in the light of the theory of fiber bundles. He observed that the applied covariant derivatives bear a very close relationship to the covariant derivatives of general relativity theory; and the quantities $F_\alpha^{(2)\mu\nu}$ are in close analogy to the curvature tensor of general relativity. Since the original YM theory derived $F_\alpha^{(2)\mu\nu}$ from a specific form for the Lagrangian density, they could not state anything more than an observed similarity. Furthermore, the Lagrangian-invariant introduction of $F_\alpha^{(2)\mu\nu}$ and their covariant derivatives also leaves free the opportunity for application to gravitational fields. An advantage of this treatment is to find conserved Noether currents which are of an identical-form in different gauge fields.

[12] I.e., which serves as the source for the gauge-fields, and consequently as the source for the characteristic charges of the given fields.

[13] Here the only condition assumed was that the field equations be satisfied. No restriction was imposed on the form of the Lagrangian density except that it be invariant under local gauge transformations as defined for the infinitesimal transformations $\delta L$.

There is easy to see that $F_\alpha^{(1)\mu\nu}(\dot{x})$ and $F_\alpha^{(2)\mu\nu}(x)$ transform in the same way, as isovectors, under the local transformation $V(\dot{x}) \in G$:

$$F_\alpha^{(1)'\mu\nu}(\dot{x}) = V^{-1} F_\alpha^{(1)\mu\nu}(\dot{x}) V \quad \text{and} \quad F_\alpha^{(2)'\mu\nu}(x) = V^{-1} F_\alpha^{(2)\mu\nu}(x) V.$$

Notice, that the forms of $J_\alpha^{(1)\nu}(\dot{x})$ conserved currents in the presence of velocity depending fields coincide with the form of those currents that we had for space-time depending fields. With respect to this identical form, as well as to the variety of the symmetry groups that they may obey, one can replace $\varphi(\dot{x}) \to \varphi(x)$, $D(\dot{x}) \to B(x)$ and $J_\alpha^{(1)}(\dot{x}) \to j_\alpha^{(1)}(x)$, where $B(x)$ are familiar physical gauge fields with symmetries, e.g., U(1), SU(2), [and SU(2)×U(1)], SU(3) or SO(3,1), with the substitution of ג by the corresponding coupling constants. $F_\alpha^{(1)\mu\nu}(\dot{x})$ take the same forms and transform in a velocity dependent **D** gauge field like the components of a $j^\nu(x)$ current and isovectors $f^{\mu\nu}(x)$ of a general matter field $\varphi(x)$ and gauge field **B**, defined by $f^{\mu\nu} = \partial^\nu B_\mu - \partial^\mu B_\nu - גB_\mu \times B_\nu$ in the four dimensional space-time. (This yields the information, that in a boundary situation, i.e., in the absence of relativistic accelerations, our derivation produces the same result as it was known without the assumption of a velocity dependent gauge field. We got back to the results that were known in the absence of a velocity-dependent gauge field, and that were based on calculations in an only space-time dependent gauge field. So, without employing accelerations, we derived the same conserved currents. This justifies our preliminary assumption, that handling the space-time coordinates as implicit parameters not only provides additional information but it preserves the physical relevance of the theory.)

*4.2.1 First conserved quantity: Conservation of the field charge (ך)*

In a general case, the $T_\gamma$ (which appear in the currents) as introduced above, are matrix-representation operators generating the group G, with the mentioned commutation rule $\left[T_\alpha, T_\beta\right] = iC_{\alpha\beta}^\gamma T_\gamma$. They can be replaced by concrete operators of the concerned fields, according to their characteristic symmetry groups, like U(1), SU(2), SU(3) or SO(3,1), and ג can be substituted by the concrete coupling constants of the individual physical fields. Thus, in a general case, and with group G of an arbitrarily chosen physical field **B**, one can write $\varphi(x)$ and **B** in the equations for the currents $J_\alpha^{(1)\mu}$ and substitute the above equations with:

$$J_\alpha^{(1)\nu}(x) = iג \frac{\partial L}{\partial(\partial_\nu \varphi_k)} (T_\alpha)_{kl} \varphi_l(x), \qquad J_\alpha^{(1)\nu}(x) = \partial_\mu F_\alpha^{(1)\mu\nu}(x),$$

$$F_\alpha^{(1)\mu\nu}(x) = \frac{\partial L}{\partial(\partial_\mu B_{\nu,\alpha}(x))}, \quad \text{and} \tag{9}$$

$$\hat{\partial}_\mu F_\alpha^{(1)\mu\nu}(x) = \partial_\mu F_\alpha^{(1)\mu\nu}(x) + iגC_{\alpha\beta}^\gamma B_{\mu,\beta}(x) \times F_\gamma^{(1)\mu\nu}(x)$$

The operators of the quanta of the given physical field are determined by the generators $\{T_\alpha\}$ of the symmetry group of the respective field. The full conserved field charge currents $J_\alpha^{(1)\mu}$ will provide the conserved quantities of the field $\varphi(x)$, which the gauge field **B** interacts with. We called these conserved quantities field charges and denoted by ך. We can get the conserved quantity by integration of the current in the usual way, applying Gauss' theorem, where the integral of the spatial components vanishes at an infinite boundary, and we get:

$$\frac{d}{dt} \frac{ג}{c} \int \frac{\partial L}{\partial(\partial_4 \varphi_k)} (T_\alpha)_{kl} \varphi_l(x) dV = 0$$

where the integral provides the conserved field charge ך of the source field $\varphi$.

The results derived in this subsection coincide with the well known conservation laws of field theories. We treat it here in order to make it comparable with the results of the next session, and to demonstrate that the two conserved quantities appear simultaneously (Sec. 4.3.3).

*4.2.2 Second conserved quantity: Conservation of the isotopic field charge spin (Δ)*

$J_\alpha^{(2)\nu}(x)$ are the isotopic field charge spin currents, which are – similar to $J_\alpha^{(1)\mu}$ – also conserved and yield a conservation law. The conserved quantity derived from $J_\alpha^{(2)\nu}(x)$ is the isotopic field charge spin $\varDelta$.

The conserved current in the kinetic field can be read from the equation

$$J_\alpha^{(2)\nu}(x) = i\lambda \left[ \frac{\partial L}{\partial(\partial_\mu \varphi_k)} (T_\alpha)_{kl} \varphi_l(\dot{x}) \lambda_\mu^\nu - C_{\alpha\beta}^\gamma D_{\dot{\omega},\beta}(\dot{x}) \lambda_\mu^\omega \times F_\gamma^{(2)\mu\nu}(x) \right]. \quad (10)$$

The right side of (10) represents the full conserved isotopic field charge spin current, which includes the contribution of the **D** field.[14]

We have introduced the **D** field – which is shown to be responsible for the isotopic field charge spin transformation – to counteract the dependence of the *V* transformation on $\dot{x}_\mu$. The field equations, which are satisfied by the twelve independent components of the **D** field, and their interaction with any field that carries isotopic field charge spin are unambiguously determined by the defined currents and covariant $F^{(2)\mu\nu}$-s constructed from the components of **D**. Considering a general Lorentz- and gauge invariant Lagrangian, we obtain from the equations of motion that $J^{(2)1,2,3}$ and $J^{(2)4}$ are, respectively, the isotopic field charge spin current density and isotopic field charge spin (*Δ*) density of the system. The total isotopic field charge spin

$$\Delta = \frac{i}{\lambda} \int J^{(2)4} d^3 x$$

is independent of time and independent of Lorentz transformation. $J^{(2)\mu}$ does not transform as a vector, while *Δ* transforms as a vector under rotations in the isotopic field charge spin field.

*4.2.3 Coupling of the two conserved quantities (ℸ and Δ)*

The dependence of the two currents $J_\alpha^{(1)\mu}$ and $J_\alpha^{(2)\mu}$ on each other has physical consequences. Once, it justifies that the quantities, whose conservation they represent and which are coupled (by $\lambda_\mu^\nu = \partial_\mu \dot{x}^\nu$), exist simultaneously. Secondly, the coupling of a conserved quantity in a space-time dependent field – which coincides with one of our familiar physical fields – with another in a kinetic (velocity dependent, new) gauge field indicates that *the derived conservation verifies* just *the invariance between two isotopic states of the field charges, namely between the potential ℸ$_V$ and the kinetic ℸ$_T$* what we intended to prove. Remember that ℸ can be field charges of different physical fields marked in common with **B**, while *Δ* represents a single quantity belonging to the kinetic gauge field **D**.

We obtained, that *in the presence of kinetic fields we have two conserved currents that are effective simultaneously*. The kinetic gauge field **D** is present simultaneously with the interacting matter [*φ*] and gauge [**B**] fields. The presence of **D** corresponds to the property of the field charges ℸ of the individual fields so that they split in two isotopic states, and analogously to the isotopic spin, we named these two states *isotopic field charge spin* what we denote by $\Delta$. The source of the isotopic field charge spin ($\Delta$) is the field $\varphi(\dot{x})$, in interaction with the kinetic gauge field **D**.

---

[14] Similar attempts (like our in the velocity space) were made by (Pons, Salisbury and Shepley, 1999) in the phase space (with a particular mapping from the configuration space to phase space), and they anticipated the quantization of the models.

The physical meaning of $\Delta$ can be understood, when we specify the transformation group associated with the **D** field, which describes the transformations of ℸ (i.e., the isotopic field charges). ℸ can take two (potential and kinetic) isotopic states ℸ$_V$ and ℸ$_T$ in a simple unitary abstract space. Their symmetry group is $SU(2)$, that can be represented by 2×2 $T_\alpha$ matrices. There are three independent $T_\alpha$ that may transform into each other, following the rule $\left[T_\alpha, T_\beta\right] = iC_{\alpha\beta}^\gamma T_\gamma$, where the structure constants can take the values 0, ±1. Let $T_1$ and $T_2$ be those which do not commute with $T_3$; they generate transformations that mix the different values of $T_3$, while this "third" component's eigenvalues represent the members of a $\Delta$ doublet. For the isotopic field charges compose a ℸ doublet of ℸ$_V$ and ℸ$_T$, the field's wave function can be written as

$$\psi = \begin{pmatrix} \psi_T \\ \psi_V \end{pmatrix}. \quad (11)$$

(11) is the wave function for a single particle which may be in the "potential state", with amplitude $\psi_V$, or in the "kinetic state", with amplitude $\psi_T$. $\psi$ in (11) represents a mixture of the potential and kinetic states of the ℸ, and there are $T_\alpha$ that govern the mixing of the components $\psi_V$ and $\psi_T$ in the transformation. $T_\alpha$ ($\alpha$ = 1, 2, 3) are representations of operators which can be taken as the three components of the isotopic field charge spin, $\Delta_1$, $\Delta_2$, $\Delta_3$ that follow the same (non-Abelian) commutation rules as do the $T_\alpha$ matrices, $[\Delta_1, \Delta_2] = i\Delta_3$, etc. These operators represent the charges of the isotopic field charge spin space, and $\psi$ are the fields on which the operators of the gauge fields act.

The quanta of the **D** field should carry isotopic field charge spin $\Delta$. The $\Delta$ doublet, as a conserved quantity, is related to the two isotopic states of field charges (ℸ), and the associated operators ($\Delta_i$) induce transitions from one member of the doublet to the other.

*4.2.4 Interpretation of the isotopic field charge spin conservation*

Invariance between ℸ$_V$ and ℸ$_T$ means that they can substitute for each other arbitrarily in the interaction between field charges of any given fundamental physical interaction. They appear at a probability between [0, 1] in a mixture of states in the wave function given by (11), so that the Hamiltonian of a *single particle* oscillates between $V$ and $T$, while the Hamiltonian of a *composite system* is a mixture of the oscillating components of the particles that constitute the system. The individual particles in a *two-particle system* are either in the $V$ or in the $T$ state respectively, and switch between the two roles permanently; while the observable value of $H$ is the expected value of the mixture of the actual states of the two, always opposite state particles.

The invariance between ℸ$_V$ and ℸ$_T$ (what is ensured by the conservtion of $\Delta$), and their abilitiy to swap means also that they can restore the symmetry in the physical equations which was lost when we replaced the general ℸ (namely mass $m$, electric charge $q$, ... etc.,) by their isotopes ℸ$_V$ and ℸ$_T$.[15]

*4.2.5 Prediction of the gauge quanta of the isotopic field charge field*

Almost all what has been mentioned afore are in accordance with our physical experience but the predicted quanta of the gauge field **D**. When we introduced the distinguished two kinds of isotopic field charges in our physical equations we reached a limit: we have got equations, where certain symmetries of the traditional equations were distorted. That was not in harmony with our experience (that means, was not in accordance with the facts). Then, we referred to a conservation law (Darvas, 2009) for a newly introduced quantity, namely the isotopic field charge spin ($\Delta$), and its invariance restored the lost symmetry.

---

[15] Consequences of the application of effective field theories were analysed e.g., in physics by S. Weinberg (1997) and in philosophy by E. Castellani (2001).

The isotopic field charge spin should exist in an above defined gauge field **D**. Such a gauge field must have quanta that carry the isotopic field charge spin $\Delta$.[16] Exchange of these quanta should mediate between the interacting field charges, so that switch the emitter field charge ר$_V$ → to ר$_T$, and the recipient field charge from ר$_T$ → to ר$_V$ and *vice versa*. These quanta have not been observed. Therefore, they do not belong to the class of our observed quantities. The test of the derived theory is whether one can find such mediating bosons. *This is the only issue in this work, where we left the field of reinterpretation of known facts, and moved to the domain of prediction*.

We will denote the *predicted* quanta of the **D** field by $\delta$. We will call this hypothetical boson "*dion*", after the Greek term meaning 'flee', 'flight', 'rout' in English. The $\delta$ quanta (dions) carry the $\Delta$ (isotopic field charge spin as a physical property: charge of the **D** field). For all the quantum numbers of the interacting field charges are mediated by the respective SM mediating bosons, the known quantum numbers of a dion ($\delta$) are 0 but the $\Delta$ – with a notice in the next paragraph. We can agree in a free convention for the sign of $\Delta$ to be + or – (½) according to whether it switches the isotopic field charge spin from a potential state to a kinetic or back.

We have to add a notice about the mass and electric charge of the dion. The mass of a mediating boson in the SM depends on the range and strength of the interaction. The masses of the gauge quanta of the individual interacting fields are in accordance with this requirement: they carry that mass. What can we expect from another boson that exists in a simultaneously present gauge field, and appears parallel (or antiparallel) with the known standard exchange bosons? Note that observing from a given reference frame, the field charges in potential state ($m_V$, $q_V$) are *bare quantities*, while the kinetic ones owe '*dressed*' or *renormalized* mass and electric charge in the same frame. When an isotopic field charge spin invariance mechanism flip-flops the value of $\Delta$ to the opposite, the mediating boson must carry the difference there and back, if the observer insists on his chosen reference frame. We have no reason to give up this physicist observer position. Therefore, the dion is supposed to owe this mass and electric charge difference.

Another strange peculiarity of the dion is that it switches ר$_V$ and ר$_T$ between scalar and vector quantities in their respective fields.

*4.2.6 Mechanism of $\Delta$ interaction*

An interaction between field charges ר can be described in the SM by a following Feynman diagram:

---

[16] Jackiw and Rebbi (1976) (with assistance by t'Hooft), invstigating the YM pseudoparticle solution by Belavin, Polyakov, Schwartz and Tyupkin (1975), which was found to be O(5) invariant, and after having applied the solution to the Dirac equation, they demonstrated that the pseudoparticle was distinguished by possessing a large *kinematical invariance group* "possibly important in future developments of the theory", although they did not analyse these kinematic consequences for the Dirac equation.

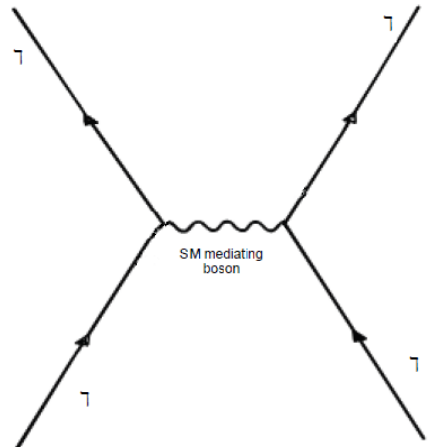

Here ꟼ can mark the field charge of any physical interaction field. In the presence of a kinetic field **D**, the interacting particles have two properties present simultaneously: ꟼ and $\Delta$, and they exchange two intermediate bosons simultaneously: the one, characteristic to the given physical interaction (photon, weak current bosons, gluons, and at any probability graviton), and in accordance with the flip of the isotopic field charge spin they exchange simultaneously δ as well. [17]

Interaction with the exchange of intermediate bosons was characterised in the Standard Model like two children throw a ball to each other with a high frequency: as soon as they receive the ball they pass it back to the other. In the model proposed in this work the two children play the same game with two balls.

The interaction could be described with a following Feynman diagram:

---

[17] Feynman (1949, Appendix D, Figure 8a and 8b) predicted the possibility of the contribution of two virtual quanta to the interaction that can be applied in our case. Goldstone, Salam and Weinberg (1962, p. 970) found also an option for two mediating bosons. They concluded – among other options – that "if part of the loss of symmetry is due to the choice of a nonivariant boson mass ..., then there must appear a two-boson pole at zero mass in the propagator of $\Phi^2$." In their conclusions Goldstone anticipated the possibility that symmetry breaking may involve "an inextricable combination of gauge and space-time transformations". It is something similar (but not the same how they later attempted) what we are trying to develop in this work. Then they noticed that Weinberg "has developed a method of rewriting any Lagrangian in order to introduce fields for bound as well as 'elementary' particles." This latter distinction impressed encouraging for my present work.

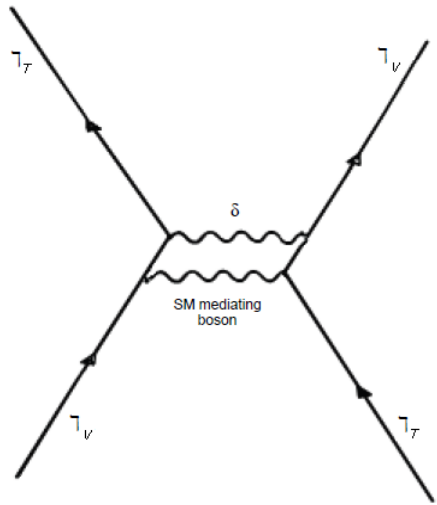

This mechanism introduces a loop in the chart what was unusual in the Standard Model. (Cf., the 2008 preliminary results of experiments at the CERN on virtual-particle quantum loop effects, e.g., The Bell collaboration, 2008; Gershon, 2008; Wilczek, 2008; CERN workshop, 2008a, Sec. 5; CERN workshop, 2008b; CERN workshop, 2008c, Secs. 4-5.) Nevertheless, making distinction between potential and kinetic states, which are in accordance with observed facts, as well as the introduction of the corresponding kinetic gauge field were unusual too. Experience must decide on the validity of the proposed model. *The criterion can be the detection of bosons to be identified as dions* ($\delta$). Maybe virtual bosons assumed to mediate in box- or penguin-loop effects, or other virtual bosons to be observed later could be identified with dions. Such identification depends on a prerequisite to replace the SUSY with the IFCS assumption in the supporting theory (cf. the Introduction).

All the rest in the isotopic field charge spin ($\Delta$) model is in accordance with experience. They are only reinterpreted. This revision predicts the existence of an invariance (of the property $\Delta$) interpreted in a kinetic gauge field (**D**) whose quanta ($\delta$) are assumed to mediate between the potential ($\psi_V$) and kinetic ($\psi_T$) states of the respective field charges ($\daleth$). We refer to this mechanism as the *Isotopic Field Charge Spin (IFCS, or $\Delta$) assumption*.

## 5. DISCUSSION

The two basic parts of a Hamiltonian can be characterised in the proposed model with the following dual properties:

| property | $H_V$ | $H_T$ |
| --- | --- | --- |
| energy | potential | kinetic |
| field | matter field | gauge field |
| localisation | in space-time | in velocity space |
| sources | field charges | field charge currents |
| transformation of the source | scalar | vector |
| isotopic field charge | $\daleth_V$ | $\daleth_T$ |
| state of the source | bound | free |
| observable status | corpuscle | wave |
| wave function component | longitudinal | transversal |

According to the afore proposed model:

(1) Nature makes a distinction between the potential ($\psi_V$) and kinetic ($\psi_T$) states of particles. Two particles, which do not differ in any other property can be distinguished whether they are in potential or kinetic states.

(2) A single particle permanently changes its state from potential ($\psi_V$) to kinetic ($\psi_T$) and back (Cf. Section 4.2.4).

(3) Interaction takes place between potential state of one particle ($ק_V$) and the kinetic state of the other ($ק_T$), and *vice versa* (Cf. Section 4.2.5).

(4) Interaction occurs by the exchange of two bosons: one of them was described in the SM and responds for the respective field charge ($ק$), (the quantum derived in Section 4.1, as a consequence of the respective conservation law and the related dynamical field equations); while the other one, the dion ($\delta$) is responsible for the swap of the isotopic field charge spin ($\Delta$), (the quantum derived in Section 4.2, as a consequence of the respective conservation law and the related field equations and equations of motion). Dion mediates between a scalar and a vector state, its only quantum number is $\Delta$ which can take two, $\pm\frac{1}{2}$ positions.

(5) There is assumed a gauge invariance, mathematically derived in (Darvas, 2009), which makes distinction between the two states. This gauge invariance is interpreted in an Isotopic Field Charge Spin ($\Delta$) gauge field (**D**), as rotations of the $\Delta$ vectors. It should characterise a symmetry of nature which was not described before.

(6) The field charge ($ק$) is a quantity conserved under the isotopic field-charge transformation (cf., Section 4.2.3).

(7) Isotopic Field Charge Spin ($\Delta$) invariance (cf., Section 4.2.3) concerns all types of particles, which can occur in both potential ($\psi_V$) and kinetic ($\psi_T$) states. (All baryons, leptons and unit mass are assumed to do so.) $\Delta$ appears as a property characterising the agents of different individual physical interactions.

This statement needs additional remarks concerning the individual fundamental physical interactions. We present examples for each interaction.

(7a) The model can be applied to the gravitational interaction with the reservation that the unit mass, as the source of the gravitational field, has not been identified, and there is also questionable its fermionic character since the graviton, as its mediating boson has spin 2. The latter assumption maintains restriction for the applicable form of the Lagrangian, but does not question the applicability of the described model. Nevertheless, gravitational mass can be naturally associated with the potential state of mass, and inertial mass can be associated in the same way with the kinetic state of mass. This interpretation provides a specific insight into the centuries old problem of the distinction and equivalence of the two masses. Graviton is responsible for the interaction between two mass units, in the course of which the two actors are in two different initial states, and $\delta$ is responsible to switch both of them in the opposite $\Delta$ state as a result of the interaction. The process, with the exchange of the two bosons, continues permanently.

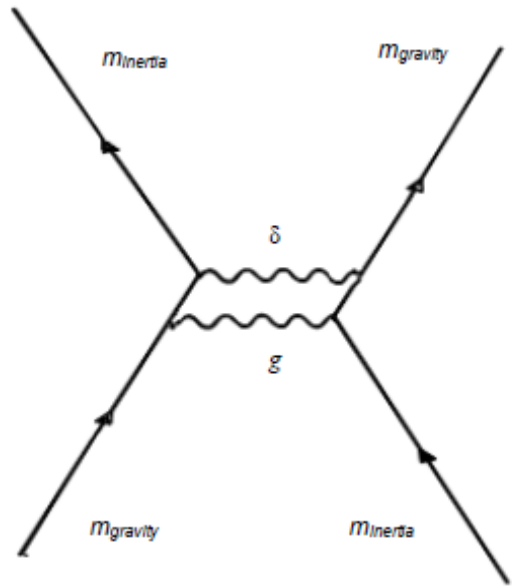

(7b) The model fits best to the electromagnetic interaction. Similar to the mass, two (potential and kinetic) states of the electric charges can be interpreted, and they correspond to the Coulomb charge and to the components of the charge current (Lorentz charges). There exists a similar equivalence principle between them, like between the two states of masses. Since the electromagnetic theory is subject to an invariance under $U(1)$ group's symmetry transformations, there belongs to it a single gauge quantum ($\gamma$). It mediates interaction between particles with potential (Coulomb) state and kinetic (Lorentz) state charges. $\delta$ is responsible to switch both actors in the opposite $\Delta$ state after the interaction. The process, with the exchange of the two bosons, continues permanently. This model corresponds to the asymmetric model described by Møller (1931). Taking into consideration the difference between the potential ($\psi_V$) and kinetic ($\psi_T$) states of ܐ (where in this concrete case ܐ denotes the electric charge) one does not need the artificial symmetrisation introduced by Bethe and Fermi (1932), because the considered asymmetry is compensated by the $\Delta$ invariance mediated by $\delta$.

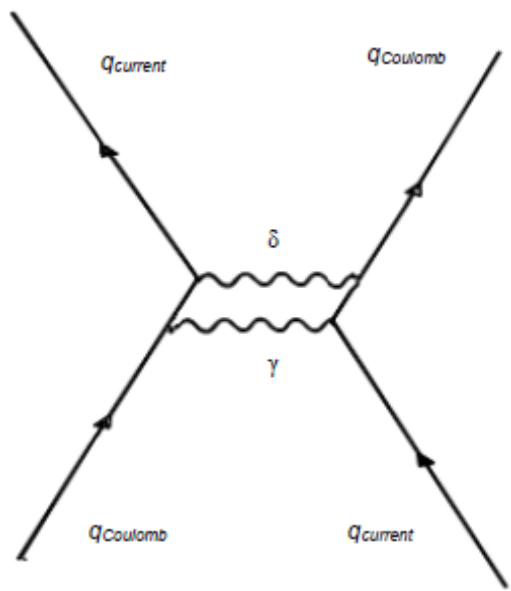

(7c) In the case of the weak interaction (coupled with the electromagnetic) the electric charge appears as one of a four-member multiplet. (Muon and tau play the same role like the electric charge.) The three other agents are flavours, as sources of the weak interaction, mediated by the three weak current bosons. The four mediating bosons are the consequence of the $SU(2) \times U(1)$ symmetry. The combined symmetry means that the agents form multiplets individually (although as members of the combined group). Thus the electron, muon and tau form doublets with their respective neutrinos, and hadrons form multiplets with a quantum mixture of their other flavoured brothers. In the framework of the Isotopic Field Charge Spin model all members of these multiplets have got a twin pair, so that if they were considered in potential state (sources of their respective field) in the SM, their twin pair has a kinetic state field charge (kinetic flavour), and the same holds for the opposite configuration. Switching between the twin pairs, that means, between the two opposite $\Delta$ states of the individual flavours during the interaction is mediated by a corresponding $\delta$. The weak current bosons and $\delta$ act (anti)parallel to $Z^0$, $W^+$ and $W^-$, like in the above discussed two types of interactions.

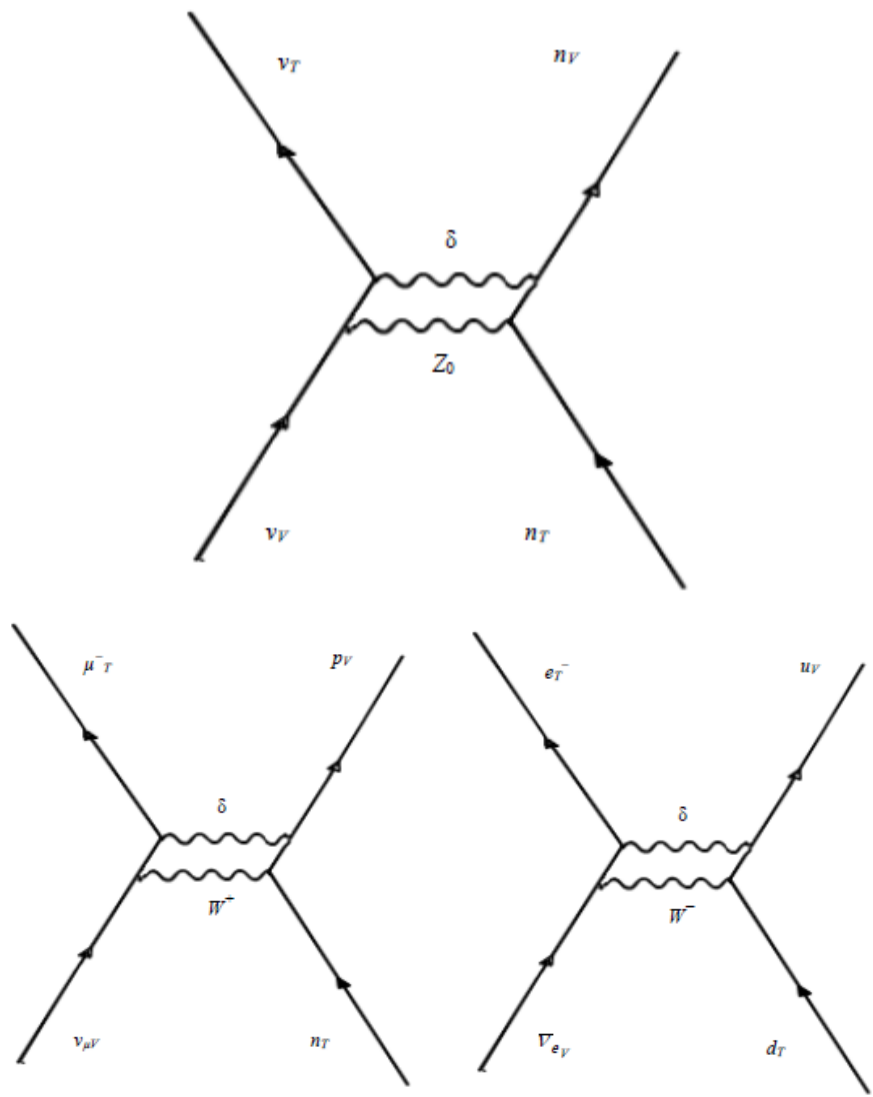

Note that the agents of weak interaction have not only weak charges, and are subjects not only of weak interaction. The role of the electric charge in electromagnetic interaction was discussed in (7b). Flavoured hadrons are also coloured and accordingly may interact strongly (7d).

(7d) Field charge of the strong interaction is colour. We assume that colour appears also in potential and kinetic states. However, due to their confinement, at least in everyday experimental circumstances, we cannot observe them in kinetic states, what supposes free motion. And yet, on the one side, we can imagine that coloured particles may oscillate within their cage limited by confinement; on the other side, they can be freed at extreme energies. Since they cannot be assumed as static objects, they must have kinetic state too. Therefore, similar to the mass, electric charges, weak charges, colour should also have $\Delta$, and appear both in potential and kinetic states. The exchange of $\delta$ bosons should take place in a similar way like the exchange of gluons is interpreted, parallel (or more probably antiparallel) with those. (See illustration on the examples of the $\pi^+$ and the proton.)

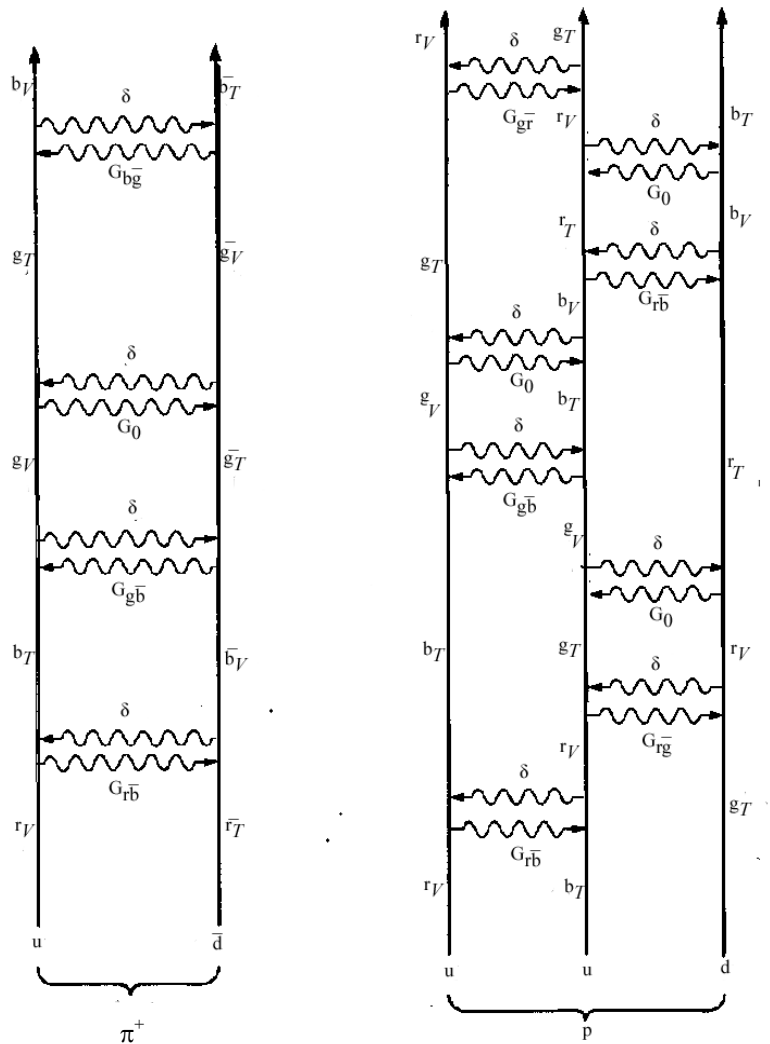

(8) Particles in potential state are bare. Renormalised mass and charge should be attributed to the kinetic states.

(9) The kinetic state particles show behaviour closer to wavicles, and potential state particles closer to corpuscles. As they commute at high frequency between the two states (at least in experiments, where they interact) they cannot be distinguished. The isotopic field charge spin assumption may provide an explanation for the double (wave–corpuscle) behaviour of massive particles. Nevertheless, the properties that determine whether being a particle or wavicle are attributed not to physical objects but to their respective individual field charges. (That means, we have no reason to assume that an object's mass, electric charge, flavour and colour are in the same isotopic *Δ* state simultaneously.)

(10) The frequent commutation between the two isotopic field charge states give no opportunity for an interacting particle to spend long time in one of the *Δ* states. This circumstance does not allow the runaway growth e.g., for an electron cloud up to infinity, whose wave function is predicted to do so in a free kinetic state, because before its wavicle (and the surrounding thundercloud) would spread out too far, it will switch to its opposite (potential) *Δ* state. When it will switch back again to kinetic state, the runaway growth and building up an electric charge thundercloud starts again from its bare appearance.[18] An analogous process is supposed to take place with other ⊣–s, like mass and colour charge in a similar way as predicted by F. Wilczek (2003, p.33). This oscillating mechanism may explain different behaviour of a fermion in certain aspects in bound and in free states.

**Acknowledgement** I express my thanks to the memory of the late *Yuval Ne'eman* for his wise advices in the first period when I developed the preliminary physical ideas and notions presented in this work during our meetings and frequent correspondence in the years 2002-2006.

---

[18] Dirac (1962, p. 64) proposed a possible model that eliminated the runaway motions for the electron.